# Atom-by-Atom Substitution of Mn in GaAs and Visualization of their Hole-Mediated Interactions

D. Kitchen[1,2], A. Richardella[1,2], J.-M. Tang[3], M. E. Flatté[3], A. Yazdani[1]

[1]*Department of Physics, Joseph Henry Laboratories, Princeton University, Princeton, New Jersey 08544, USA.* [2]*Department of Physics, University of Illinois at Urbana-Champaign, Urbana, Illinois 61801, USA.* [3]*Optical Science and Technology Center and Department of Physics and Astronomy, University of Iowa, Iowa City, Iowa 52242, USA.*

**The discovery of ferromagnetism in Mn doped GaAs[1] has ignited interest in the development of semiconductor technologies based on electron spin and has led to several proof-of-concept spintronic devices[2-4]. A major hurdle for realistic applications of $Ga_{1-x}Mn_xAs$, or other dilute magnetic semiconductors, remains their below room-temperature ferromagnetic transition temperature. Enhancing ferromagnetism in semiconductors requires understanding the mechanisms for interaction between magnetic dopants, such as Mn, and identifying the circumstances in which ferromagnetic interactions are maximized[5]. Here we report the use of a novel atom-by-atom substitution technique with the scanning tunnelling microscope (STM) to perform the first controlled atomic scale study of the interactions between isolated Mn acceptors mediated by the electronic states of GaAs. High-resolution STM measurements are used to visualize the GaAs electronic states that participate in the Mn-Mn interaction and to quantify the interaction strengths as a function of relative position and orientation. Our experimental findings, which can be explained using tight-binding model calculations, reveal a strong dependence of ferromagnetic interaction on crystallographic orientation. This anisotropic interaction can potentially be exploited by growing oriented $Ga_{1-x}Mn_xAs$ structures to enhance the ferromagnetic**

**transition temperature beyond that achieved in randomly doped samples. Our experimental methods also provide a realistic approach to create precise arrangements of single spins as coupled quantum bits for memory or information processing purposes.**

In $Ga_{1-x}Mn_xAs$, substituted Mn atoms at Ga sites act as acceptors donating holes that are believed to be responsible for mediating ferromagnetic interactions between Mn *d*-orbital core spins. Mean-field theories based on this scenario have captured the increase of ferromagnetic transition temperature, $T_c$, with doping and other macroscopic properties of $Ga_{1-x}Mn_xAs$[5,6]. Experimental efforts have also shown that improved material quality, achieved through various methods including annealing of random alloy samples, can significantly increase $T_c$ even above 150 K[7-10]. However, currently missing is an accurate microscopic picture, which identifies the precise nature of hole states involved and provides clues on how to enhance ferromagnetism in this compound. Previous STM experiments have mapped the shape of single acceptor states for Mn in subsurface sites[11], but have been unable to directly observe interactions between them[12]. First-principles calculations[13,14] have provided insight; however, their predictions for the structure of Mn-Mn interactions could only be tested statistically in macroscopic materials that also contained large numbers of defects, interstitials, and other complicating features.

We perform our experiments using a home-built cryogenic STM that operates at 4 K in ultra-high-vacuum. We use wafers of GaAs doped with $10^{18}$-$10^{19}$ Zn atoms/cm$^3$, which are cleaved *in situ* to expose a (110) surface. The cleaved samples show STM topography and spectroscopy that are characteristic of a degenerately doped p-type GaAs sample, with the Fermi level very close to the valence band (VB) edge of the GaAs[15]. Although the (110) surface undergoes a small reconstruction, there are no surface states at energies within the GaAs gap to complicate our studies[16]. To substitute

Mn atoms for Ga, we first deposit a small concentration of Mn atoms (0.1-2% monolayer) from an *in situ* source onto the cold sample surface. The deposited Mn atoms are weakly adsorbed on the surface, appearing in VB images (tip-sample bias $V<0$) of a GaAs (110) surface as small protrusions (Fig. 1a and b) and can be manipulated using the STM tip.

We have discovered that application of a voltage pulse in the STM junction, which injects energetic electrons onto a Mn adsorbate's site, results in the substitution of one Mn for one Ga atom in the surface layer. This substitution process requires injecting electrons with energies of 0.7 eV or higher. Varying the precise placement of the tip near the adsorbate results in either its substitution or its lateral motion on the surface. Fig. 1c and d, after tip-voltage pulses, reveal the presence of the substituted Mn by showing its influence on neighbouring As atoms and the ejection of the Ga atom it has replaced. The precise physical process for the STM-assisted substitution is still under investigation; nonetheless, model calculations show that the substituted configuration of Mn has a lower energy than the adsorbate configuration on the GaAs (110) surface[17]. We have found the STM-assisted incorporation and motion of Mn adsorbates can be used to successfully substitute Mn atoms at precise Ga sites and to remove the ejected Ga atoms from the substitution area. All the experimental results reported here were repeated for Mn in different locations of the surface to ensure that interaction with native Zn acceptors in the substrate did not alter our experimental findings.

The modifications of the local density of electronic states (LDOS) of GaAs due to Mn substitution significantly impact the nature of their interactions within the GaAs host. We probed these modifications by combining real space imaging together with spatially resolved STM spectroscopy, performed by measuring the differential conductance, $dI/dV$ versus $V$, using standard lock-in techniques. The spectroscopic measurements performed in the vicinity of an isolated substituted Mn are shown in Fig. 2a. The modification of the VB electronic states ($V$<0 in the spectra), as indicated by an arrow in the tunnelling spectra of Fig. 2a, is strongest on the As neighbours nearest the Mn. The spatial extent of the Mn-induced modifications of the VB states is weak beyond the nearest As neighbours and is spatially anisotropic, as shown in Fig. 2b. The Mn has a more dramatic influence, however, due to the binding of an acceptor state to the Mn site producing a strong resonance in the tunnelling spectra inside the GaAs gap (Fig. 2a). The dominance of the Mn acceptor state in the LDOS lends itself to direct mapping of the acceptor state wavefunction in the unoccupied state images. Such images, as shown in Fig. 2c, demonstrate that the Mn acceptor has an anisotropic star-shaped spatial structure that is distributed over more than a $20Å^2$ area of the surface[18]. The large energy width (>150meV) of the conductance peak shows that the Mn acceptor state has a large overlap with the continuum states, such as those due to Zn acceptors, in our substrates, eliminating the possibility of charging effects when tunnelling through this state.

The measured modifications of the LDOS, induced by Mn in the (110) surface layer of GaAs, are consistent with tight-binding calculations of the electronic states in bulk GaAs near Mn acceptors (see supplementary information).[19] Fig. 2d shows a simulated STM image of the acceptor state based on the bulk tight-binding calculations that model the experimental situation by showing the calculated electronic states near a Mn dopant in the bulk at the (110) layer containing the Mn. These calculations have essentially one free parameter, the p-d hybridization induced by the Mn site, which is adjusted to match the measured bulk acceptor level energy. The spatial structure of the calculated wavefunction for the bulk acceptor shows a similar anisotropic shape and extent as seen in the STM experiment for Mn at the surface. The tight-binding calculations also predict resonances deep within the VB[19], which account for the VB modification imaged in Fig. 2b. These resonance states have been anticipated[19] to play a role in interpretation of photoemission[20] and infrared optical conductivity[21,22] measurements of $Ga_{1-x}Mn_xAs$.

More detailed comparisons of our measurements with the bulk calculations and with other STM measurements of subsurface Mn[11] provide ways to assess the significance of the surface in our findings. Simulating the impact of the surface on the acceptor state in the bulk calculation by adjusting the potential energy of one of the As bonded to the Mn, we find that the acceptor state shifts deeper into the gap. Our experimentally measured acceptor state is indeed found deeper in the gap (850 meV), as compared to the measured value for the bulk (113 meV), although a significant part of the observed shift is due to tip-induced band-bending typical of tunnelling between a metal and semiconductor[15,16]. More importantly, contrasting our measurements with recent STM studies of MBE-grown $Ga_{1-x}Mn_xAs$ samples[11], the subsurface Mn acceptors show states with similar anisotropic symmetry and spatial extent to those reported here. In the studies of subsurface Mn, tunnelling through the layers of GaAs broadens some of the finer features of the acceptor state compared to that reported here—a behaviour captured by our tight-binding model (see supplementary information). Overall, comparison of our data to bulk calculations and to previous STM measurements of Mn in subsurface sites shows that surface effects do not significantly alter the acceptor state's characteristics. This comparison motivates the study of the interaction between STM-substituted Mn atoms as a model situation for probing the nature of Mn-Mn interactions mediated through the GaAs host.

Our atom-by-atom substitution technique provides a powerful method to implant Mn acceptors at precise relative separations and orientations in the GaAs surface and to examine their interaction in pairs. Fig. 3 details measurements of an STM-substituted Mn pair, separated by 8 Å along a <110> crystallographic direction. Fig. 3a and 3b show changes in topography induced in the GaAs VB and in-gap states by the Mn pair.

Direct evidence for interaction within the pair can be observed in the tunnelling spectra of Fig. 3c, which reveals splitting of the acceptor state into two resonances. Fig. 3d and 3e show results of conductance mapping, which probes the spatial characteristic of electronic states by visualizing changes in $dI/dV$ at specific energies, confirming that the split states observed in the spectra (Fig. 3c) indeed have bonding/anti-bonding characteristics. In this situation, the observed bonding state occurs at higher energies because of the hole-like nature of the states involved, for which the continuum is at the top of the VB. The observed splitting in Fig. 3 also indicates that the spin states associated with the Mn's 3*d* orbitals are ferromagnetically aligned. Since the formation of bonding and anti-bonding states requires electronic states that are degenerate in both energy and spin alignment, anti-ferromagnetically aligned Mn pairs would have anti-aligned acceptor states that would not show any splitting[23]. Within a doping-dependent superexchange model of magnetism (e.g. Ref. 24), this state splitting is the dominant contribution to the Mn-Mn spin interaction.

Constructing Mn pairs using our atom-by-atom substitution technique and measuring the energy splitting of their acceptor states provide a method to map the relative strength of Mn-Mn interactions as a function of distance and crystallographic orientation in GaAs. A subset of the Mn pairs we have examined is shown in Fig. 4a-4d and Fig. 3b, along with measurements of the splitting energy of the acceptor levels for Mn pair configurations up to the six nearest-neighbour in Fig. 4e. Our key observation is that the Mn-Mn interaction decays rapidly with increasing separation between the Mn acceptors and is highly anisotropic. The data in Fig. 4e clearly show that pairs constructed along <110> crystallographic directions have a much stronger splitting than those along the <100> or <111> directions. Our ability to probe the energy splitting of the pairs is limited by the energy width of a single Mn acceptor state. Consequently, based on experimental measurements alone, we cannot rule out the possibility that pairs, such as the one separated by 5.65 Å along a <100> direction, that do not show splitting, are not anti-ferromagnetically aligned (see supplementary information). Nonetheless, the experimental measurements of the relative strength of the Mn-Mn interactions and their anisotropic character can be used as a test bed for microscopic models of ferromagnetic interactions in GaAs.

As a starting point for understanding our experimental findings for Mn pairs, we turn again to bulk tight-binding calculations, which successfully accounted for the Mn-induced changes in the LDOS of GaAs (Fig. 2d). We calculate splitting of the acceptor states for two bulk Mn atoms at the five nearest-neighbour pairs probed in the STM experiments, assuming that their core spins are ferromagnetically aligned and point along a <100> direction, the bulk easy axis. As shown in Fig. 4e, the results of these tight-binding calculations remarkably agree in the overall trend and the observed anisotropy of the experimental data. The discrepancy in the actual values of acceptor level energy splitting is likely a result of the non-trivial shift of these levels due to tip-induced band-bending in the experiment. The tight-binding model can also be used to estimate the energy gained for ferromagnetic alignment and its connection to the acceptor state energy splitting (see supplementary information).

Looking beyond the tight-binding model, first principle *ab initio* calculations also predict the same anisotropy in ferromagnetic interactions[24] that we find in our STM measurements of the state splittings for the first several nearest-neighbour pairs. In fact, various theoretical models that include detailed band structure of GaAs or the effects of spin-orbit coupling predict anisotropic Mn-Mn interactions of various degrees[25,26]. Detailed examination of our experimental data for a single Mn shows that the origin of the anisotropic splittings is most likely the shape of the acceptor wavefunction, which favours resonant interaction along the <110> crystallographic directions. However, VB modification of the electronic states due to Mn has a similar anisotropy, which could also mediate anisotropic interactions with a similar symmetry. In addition, spin-orbit interactions also produce crystalline magnetic anisotropy in GaMnAs. Our tight-binding model, which can be used to estimate this anisotropy for Mn pairs, shows that their spin alignment favours the <100> direction at the surface (see supplementary section).

Besides providing experimental results for testing models of Mn-Mn interactions in GaAs, our experiments provide microscopic information that potentially can be used to enhance ferromagnetism in $Ga_{1-x}Mn_xAs$. The significance of the experiments is perhaps best realized by considering the separations between Mn acceptors in randomly doped $Ga_{1-x}Mn_xAs$ at concentrations required for the highest reported ferromagnetic temperatures[7-10]. At 5% doping, 98% of the Mn dopants have one or more neighbour within the sixth nearest-neighbour or closer. The experiments reported here suggest that randomly distributed Mn doping fails to take advantage of the large strength of the interaction along the <110> direction. MBE-grown heterostructures could be optimized to increase the concentration of Mn dopants with <110> neighbours, and potentially enhance ferromagnetic Mn-Mn coupling. In addition, the possibility that some of the closely spaced pairs, such as the one separated by 5.65 Å along a <100> direction, could be antiferromagnetically aligned raises the question of whether spin frustration[25] could be hampering efforts to increase the ferromagnetic transition temperature in the randomly doped system.

The atom-by-atom substitution technique we have demonstrated here can also be an important tool in a number of new directions. A direct extension would be to use our technique to substitute other dopants in various semiconducting surfaces, as a method to search for strongly interacting ferromagnetic donors or acceptors. Such an effort can be guided by theoretical efforts similar to those described here. Single dopant substitution also provides a controlled method to create single spin quantum bits in semiconductors[27]. The tight-binding calculations show the exciting possibility that the spin-orbit coupling can dictate the shape of a single acceptor wavefunction, hence providing a method for obtaining spin information from measurements of LDOS distribution[28].

This work was supported by US ARO MURI and US NSF.

Correspondence should be addressed to Ali Yazdani, yazdani@princeton.edu.

**Supplementary Information** accompanies the paper on **www.nature.com/nature**.

**Figure 1 Single atom substitution of one Mn for one Ga atom. a**, Large topograph of the occupied states (500x150 Å$^2$; -2 V; 0-1.5 Å) showing many Mn adsorbates and a few subsurface Zn dopants. **b-d**, High resolution topographs (75 Å$^2$; -2 V; 0-1.3 Å). **b,** Mn adsorbate and a Zn acceptor visible on the As sublattice. **c**, STM-assisted incorporation of the Mn into a Ga site forcing the substituted Ga atom to the surface. **d**, Tip-pulses move the Ga adatom from the incorporation site, isolating a Mn acceptor ($Mn_{Ga}$).

**Figure 2 High-resolution measurements of a single Mn acceptor ($Mn_{Ga}$). a**, Spatially-resolved dI/dV measurements near a Mn acceptor. The As neighbours show enhancements deep in the VB, highlighted by the blue arrow. The large in-gap resonance over the Mn is its acceptor level. **b**, 40 Å$^2$ topograph of the occupied states (-1.5 V; 0-0.4 Å) showing enhancements concentrated on the neighbouring As. **c**, 40 Å$^2$ topograph of the unoccupied states (+1.55 V; 0-4 Å) revealing the anisotropic shape of the Mn acceptor. **d,** Tight-binding model of a Mn acceptor in bulk GaAs with the Mn spin oriented in-plane along [001]. The image shows a 40 Å$^2$ area of the (110) plane containing the Mn. The log of the integrated DOS from the Fermi level to 1.55 eV is displayed with a spatial broadening factor of 1.7 Å.

**Figure 3 Mn pair substituted in Ga sites spaced at 8 Å apart in a <110> orientation. a,** 40 Å$^2$ topograph of the VB states (-1.5 V; 0-0.4 Å). **b**, 40 Å$^2$ topograph of the in-gap states showing the bound states of Mn acceptors (+1.1 V; 0-4 Å). **c**, dI/dV measurements near the Mn pair, revealing two resonance levels in the gap. **d**, Differential conductance energy map at +1.35 V displaying the bonding nature of the higher energy resonance. **e,** Energy map at +0.91 V displaying the anti-bonding nature of the lower energy peak.

**Figure 4 Topographs (40 Å$^2$, +1.5 V) of several Mn pairs and the acceptor level splitting energy of these pairs. a**, Nearest-neighbour Mn acceptor pair (4 Å spacing, <110> orientation). **b**, Second nearest-neighbour pair (5.65 Å spacing, <100> orientation). **c,** Third nearest pair (6.92 Å spacing, <211> orientation). **d**, Sixth nearest pair (9.79 Å spacing, <111> orientation). Fig. 3b shows the fourth nearest pair (8 Å spacing, <110> orientation). The fifth nearest neighbour is not accessible in a single (110) plane. **e,** Comparative plot of the Mn acceptor level splitting for Mn-Mn pair interactions. Theory and experiment both show Mn-Mn interactions are highly anisotropic, favouring <110> orientations. Error bars correspond to the standard deviation of many pairs measured on p-type GaAs (110) surfaces.

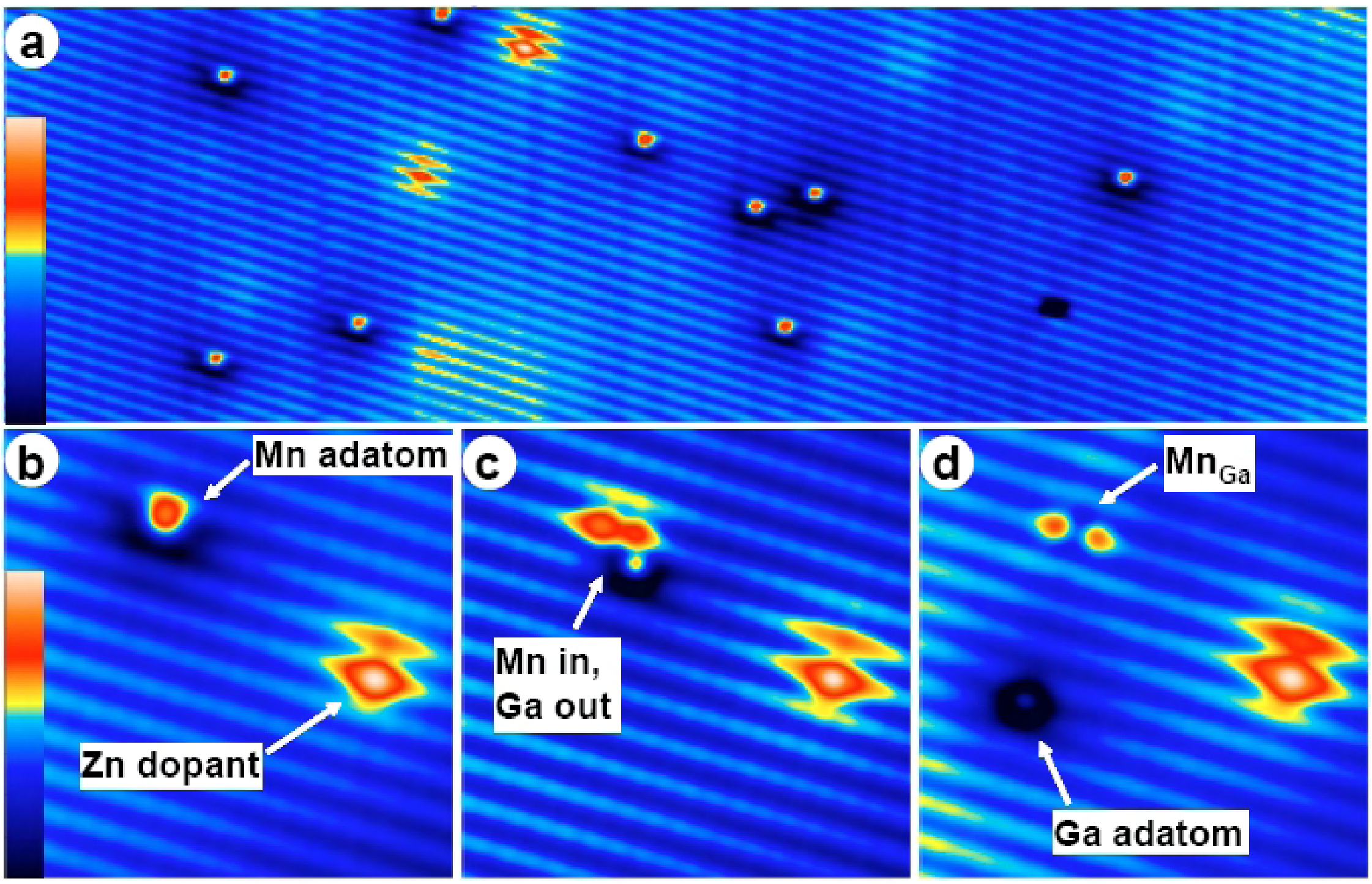

Figure 1. Kitchen *et al*.

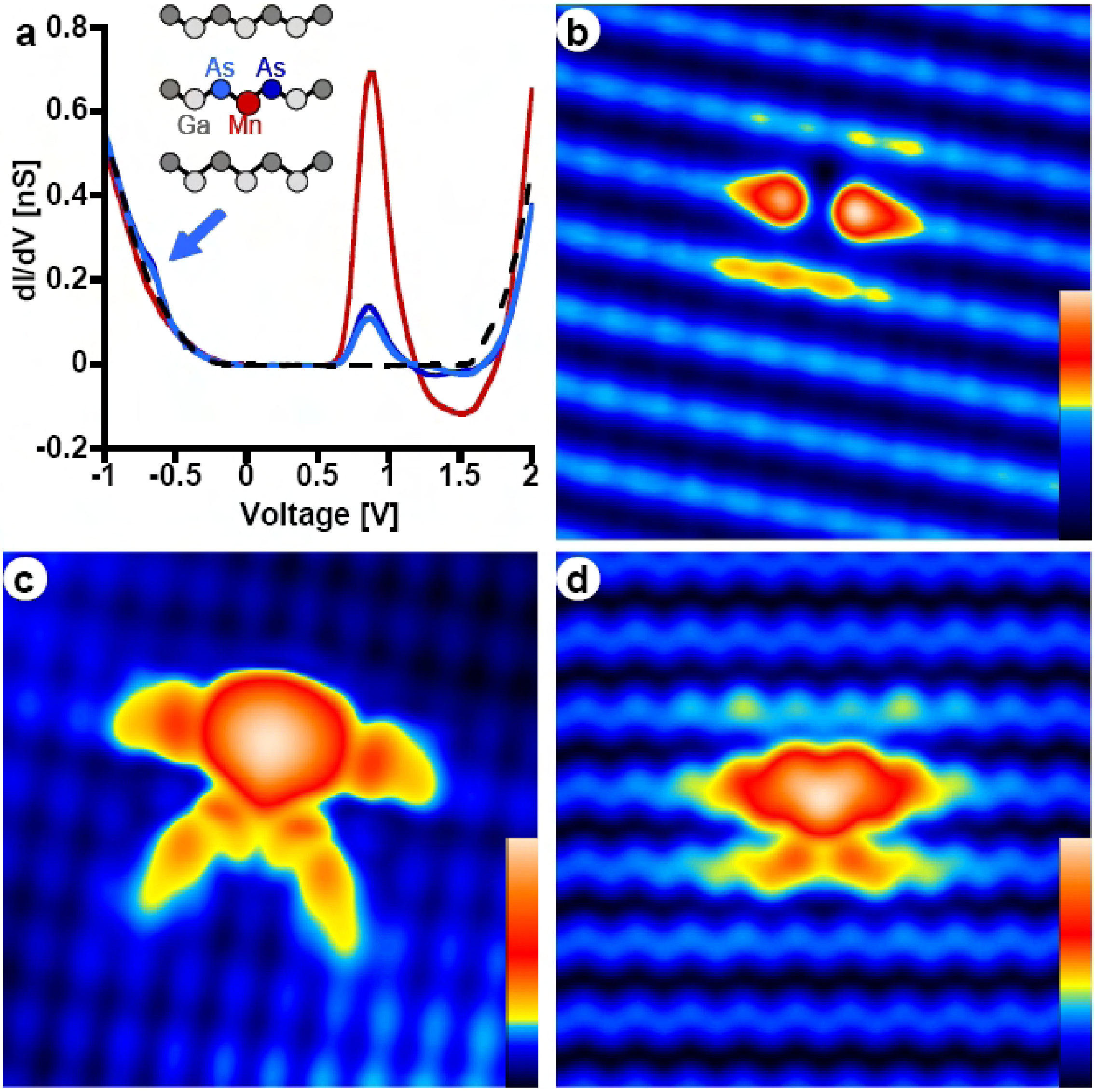

Figure 2. Kitchen *et al*.

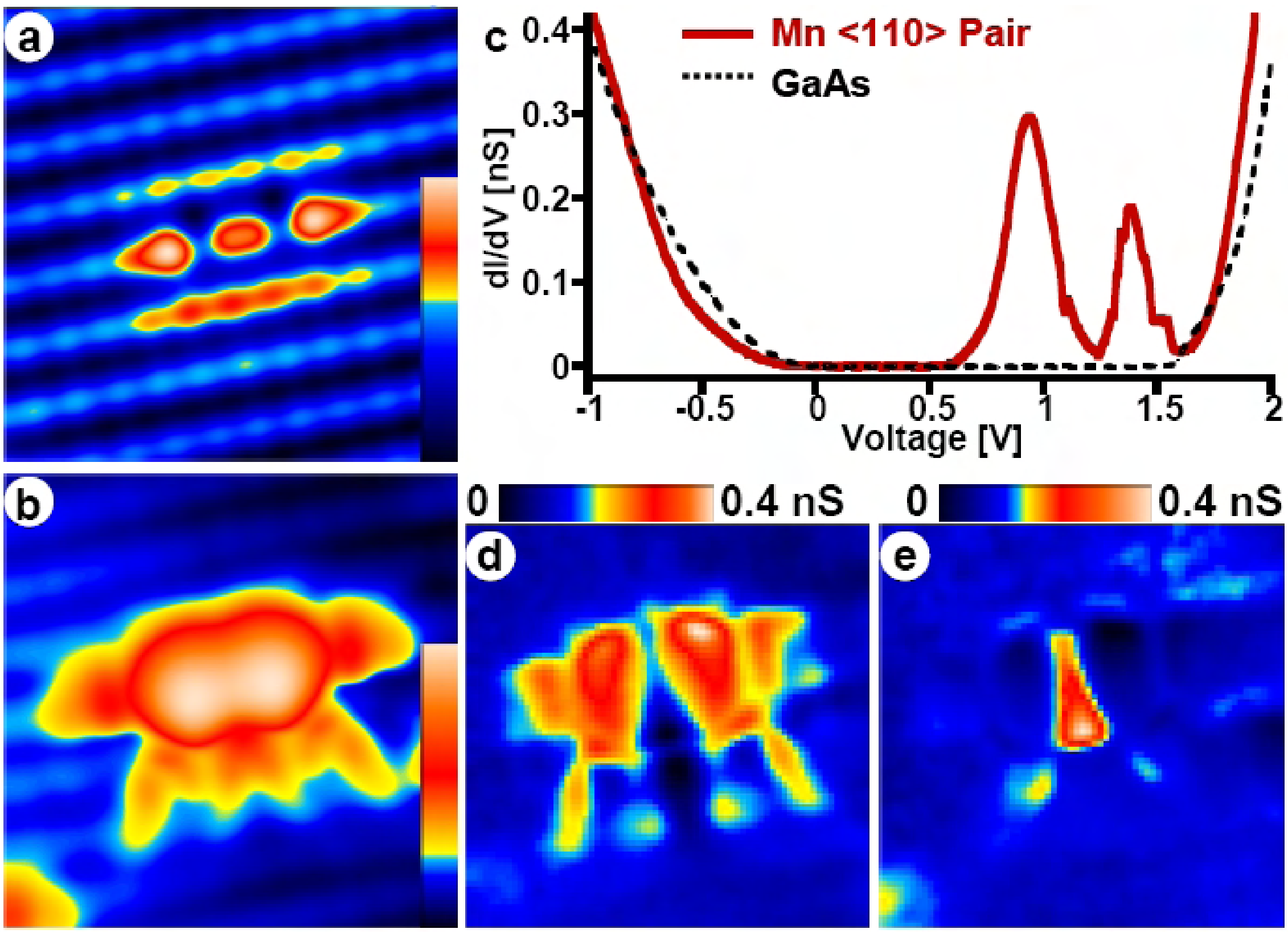

Figure 3. Kitchen *et al*.

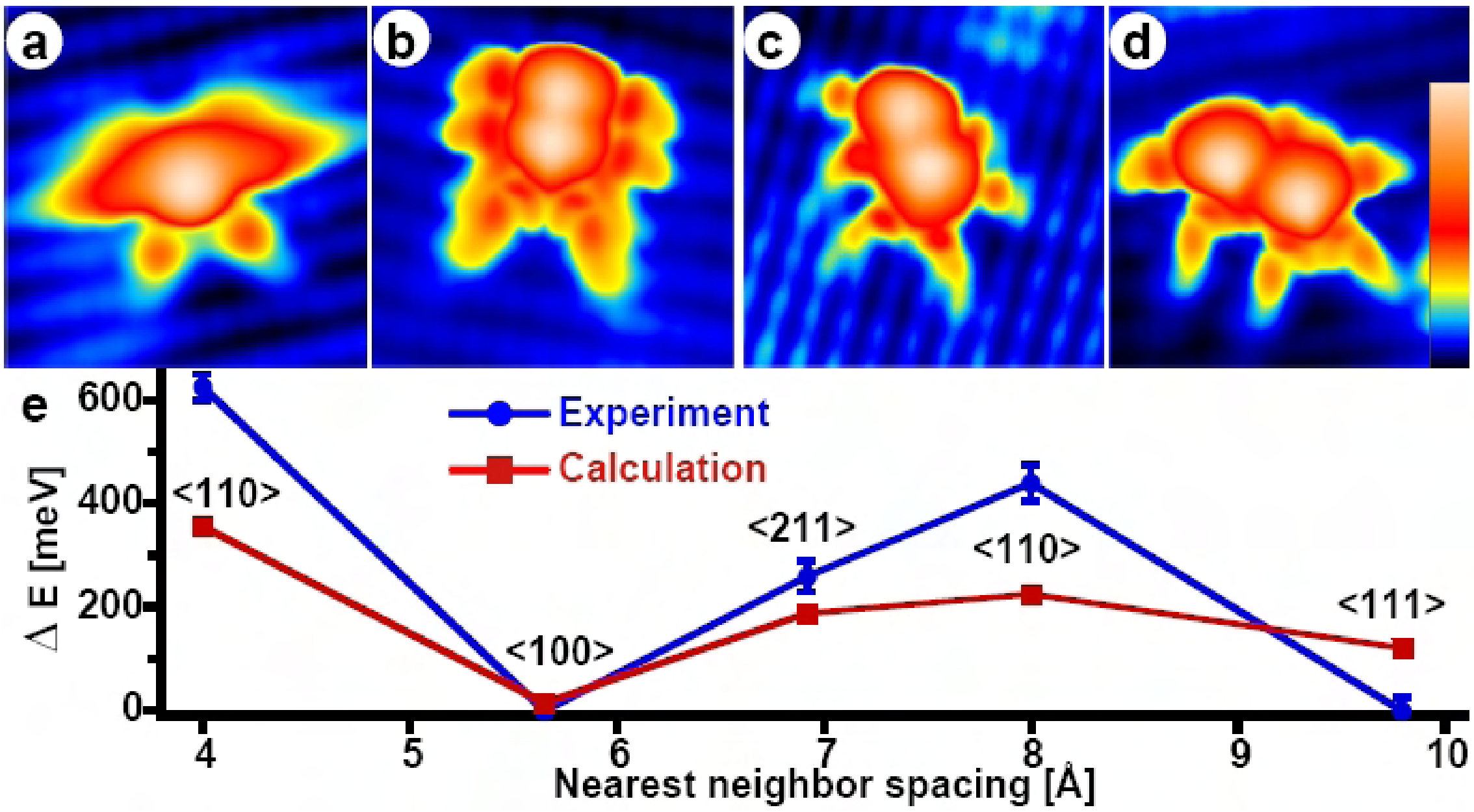

Figure 4. Kitchen *et al*.

# Supplementary Information

## Atom-by-Atom Substitution of Mn in GaAs and Visualization of their Hole-Mediated Interactions

D. Kitchen[1,2], A. Richardella[1,2], J.-M. Tang[3], M. E. Flatté[3], A. Yazdani[1]

*[1]Department of Physics, Joseph Henry Laboratories, Princeton University, Princeton, New Jersey 08544, USA. [2]Department of Physics, University of Illinois at Urbana-Champaign, Urbana, Illinois 61801, USA. [3]Optical Science and Technology Center and Department of Physics and Astronomy, University of Iowa, Iowa City, Iowa 52242, USA*

**Overview**

We report the use of a novel atom-by-atom substitution technique with the scanning tunnelling microscope (STM) to perform the first controlled atomic scale study of the interactions between isolated Mn acceptors mediated by the electronic states of GaAs. High-resolution STM measurements are used to visualize the GaAs electronic states that participate in the Mn-Mn interaction and to quantify the interaction strengths as a function of relative position and orientation. Our experimental findings, which can be explained using tight-binding model calculations, reveal a strong dependence of magnetic interaction on crystallographic orientation. The following supplementary section augments the Letter giving additional details about the tight-binding model and experimental results, along with pertinent discussions.

**Tight-binding model**

The local electronic structure around Mn atoms in GaAs is calculated using a 16-band $sp^3$-tight-binding model with spin-orbit interaction[29]. The Mn atoms are modelled with an effective potential consisting of an on-site Coulomb term of 1 eV and a spin-polarized term at the 1st nearest-neighbour As sites of 3.59 eV to describe the hybridization between the Mn d-orbitals and As p-orbitals. These values produce the correct acceptor state binding energy in bulk GaAs. The spin of the Mn 3d electrons is assumed to be aligned along a given axis. The LDOS were obtained by calculating the imaginary part of the lattice Green's functions in real space with an energy linewidth of 10 meV. Further details are available in Ref. 19 in the Letter. No additional parameters were introduced for calculating the properties of Mn pairs. For the 4 Å pair, we assume the potential is additive at one of the neighbouring As sites shared by the two Mn atoms. The splittings of the bonding and anti-bonding states are derived from the positions of the two deepest peaks in the calculated LDOS spectra in the GaAs energy gap.

**Mn near GaAs (110) surfaces**

Our tight-binding model can be used to map the LDOS near the Mn acceptor, which can then be compared to STM experiments. Supplementary figure 1 shows (110) cuts of the integrated LDOS near a Mn. Each cut captures the spatial structure of the Mn acceptor in a layer at or near the Mn, showing how the view of the anisotropic structure of a Mn acceptor evolves. The features of the layer containing the Mn (see Sup. Fig. 1a), compare well with STM measurements near the surface reported in the Letter. For views of the DOS four or five layers away from the Mn, the calculation successfully compares with previous STM measurements of subsurface Mn acceptors (see Ref. 11).

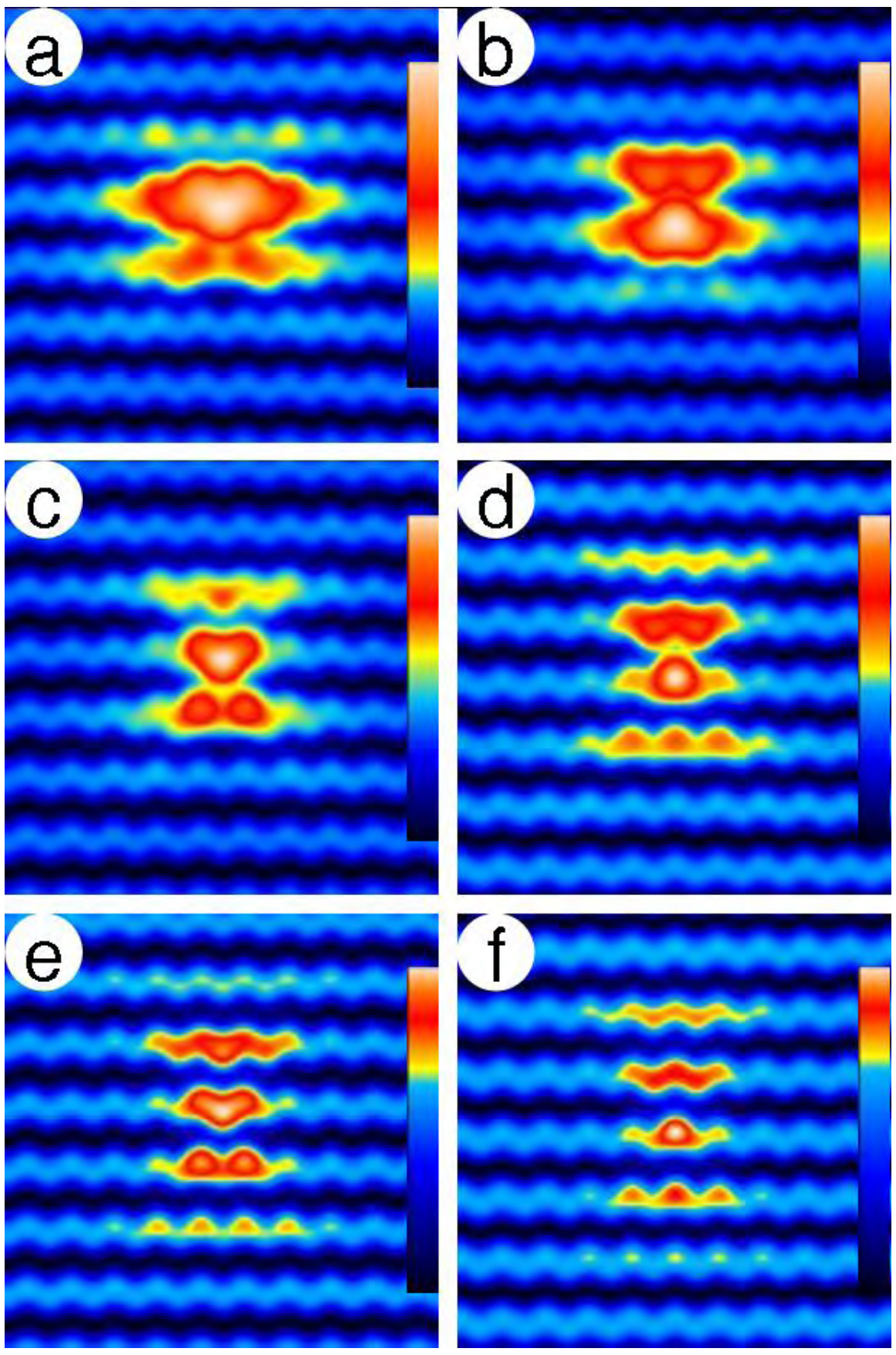

**Supplementary Figure 1 | Tight-binding model of the Density of States (DOS) near a single Mn, viewed from different (110) layers**. Each panel shows a single Mn in a 40 Å$^2$ area. The images simulate STM topography of in-gap states by integrating the calculated DOS from the Fermi energy (0 eV) to the conduction band edge (1.55 eV) for p-type GaAs. **a**, the (110) layer containing the Mn. **b-f**, (110) layers one through five away from the Mn, respectively. The calculations capture the spatial structure in STM topographs of Mn in different layers near a GaAs (110) surface.

## Mn pairs without splitting

Measurements of Mn pairs measured at 5.65 Å spacing, which are oriented across the row along a <100> direction, do not show bonding/anti-bonding behaviour. Differential conductance measurements near 5.65 Å pairs reveal no modification of the isolated acceptor level resonance. In addition the peak width of the acceptor level near 5.65 Å pairs can be compared with the isolated Mn. Measurements of many 5.65 Å pairs formed on GaAs show that the peak width measured near the 5.65 Å Mn pairs is identical to that of the isolated Mn. Supplementary figure 2 shows an example dI/dV measurement of a 5.65 Å and an isolated Mn, showing that the peak energy and peak width are experimentally identical.

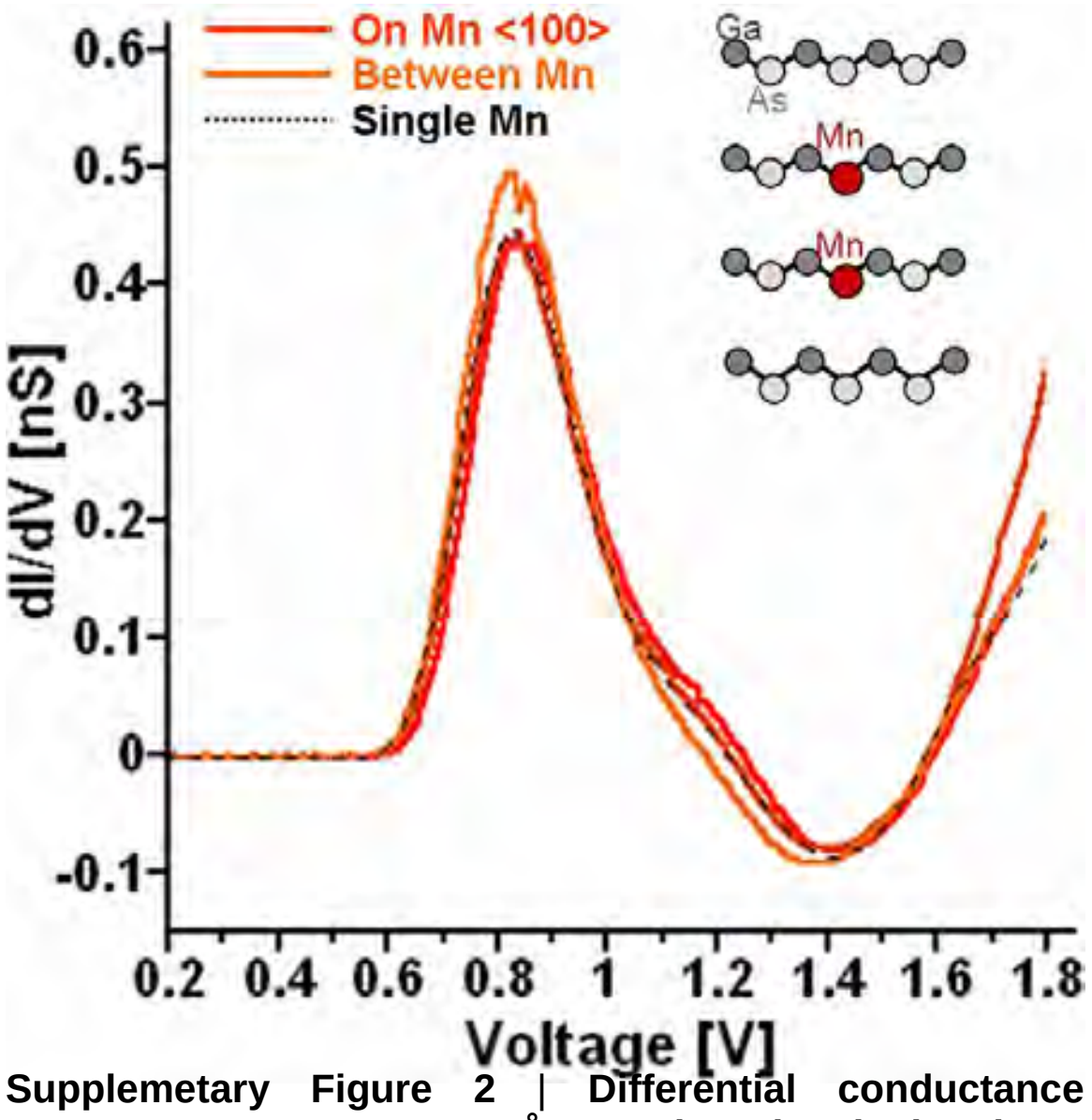

**Supplemetary Figure 2 | Differential conductance measurements near a 5.65 Å Mn pair and an isolated Mn.** The dI/dV measurements were taken with the same microtip over a 5.65 Å pair and a nearby isolated Mn. The in-gap resonances associated with the Mn acceptors have the same width and peak energy. Measurements over many Mn acceptors show that no splitting is observed or change in the peak width for the two Mn in a 5.65 Å pair configuration.

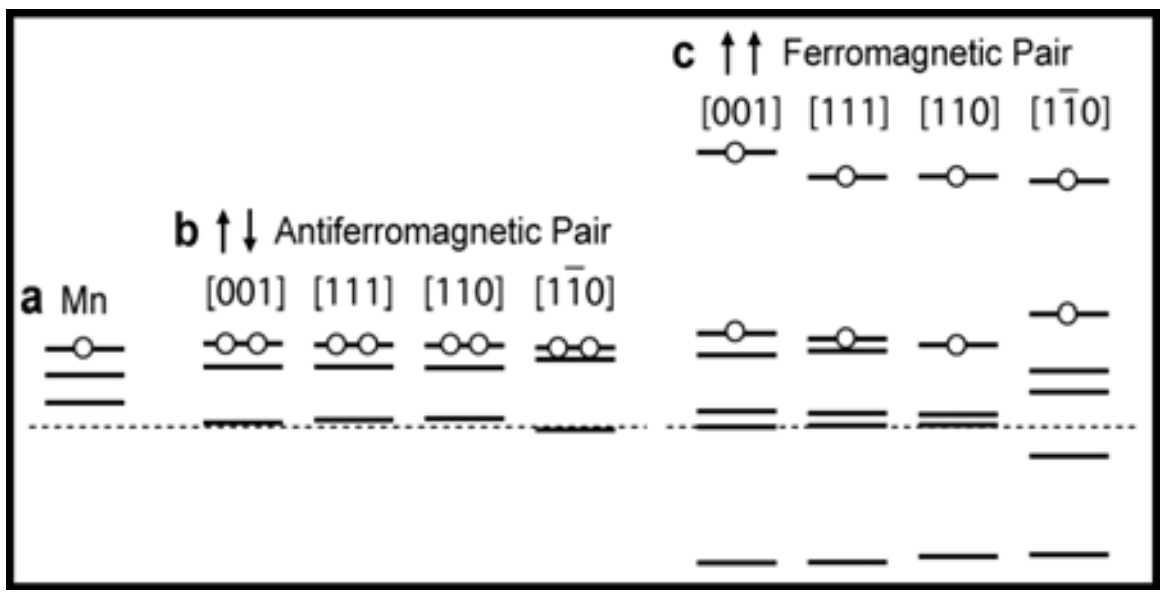

**Supplementary Figure 3 | Energy levels for a single and a pair of Mn. a**, The levels for an isolated Mn (0.113 eV, 0.075 eV, and 0.03 eV) relative to the valence band maximum (the dashed line). The levels for a Mn pair (separated by 8 Å) are shown for **b**, the antiferromagnetic alignments and **c, the** ferromagnetic alignments.

## Bonding and anti-bonding states of Mn pairs

In our tight-binding model an isolated Mn atom with a fixed core spin has three possible states for the bound hole. All three states are nearly 100% spin polarized with respect to the Mn core spin. As a result of the spin-orbit interaction in GaAs, the three states are split in energy and have different orbital angular momentum (l=1) projection along the Mn core spin axis. For a single isolated Mn only one state is occupied by a hole, and thus only one state would be seen when tunnelling in an electron (positive STM voltage). For two nearby Mn atoms these bound states can hybridize and form molecular-like states. Generally, there are six energy levels corresponding to three pairs of bonding and anti-bonding states in the ferromagnetic Mn spin configuration, and three doubly degenerate levels in the anti-ferromagnetic configuration. The double degeneracy of states for anti-aligned Mn pairs is only approximate due to the lack of inversion symmetry in GaAs, but any splitting of this degeneracy is generally not resolvable within our 10-meV resolution. Examples of these energy levels are shown in Supplementary Figure 3. Only two states are seen when tunnelling into Mn pairs at positive STM voltage, as only two states are occupied by holes.

**Energy gain for ferromagnetic interactions**

For a given pair the average energy of the six states, such as those shown in Supplementary Figure 3, is approximately the same for all spin orientations (parallel versus antiparallel and relative to the crystal axes). Thus the difference in energy between parallel and antiparallel alignment for a given pair, which is the spin-spin interaction energy J, can be estimated by adding up the energy of the four occupied states in each of the two configurations and comparing their values. As a result, pairs with large energy splitting of the two upper-most levels have lower energy in the ferromagnetic configuration. One cannot rule out by our measurements, however, the possibility of some ferromagnetic interaction for pairs with small splitting. Within this simple six-state model we have estimated J for the pairs reported in the Letter and find that the anisotropy of J is similar to the anisotropy of the splittings shown in the Fig. 4e of the Letter.

As shown in Supplementary Figure 3 the energies of the Mn pair states are also slightly modified according to the orientation of the spin relative to the crystallographic axes. From this we obtain an estimate of the crystalline anisotropy energy for each pair through the same process. The easy axis is always either along the [001] direction, which is the bulk easy axis lying in the plane of the surface, or along the [110] direction, which is parallel to the surface normal. The crystalline anisotropy we find from this is small, and varies considerably from pair to pair. As we have not taken into account surface-induced magnetic anisotropy for the pairs, which would favour the [001] direction, it appears likely that for all the pairs the [001] direction is the preferred spin orientation. In fact, assuming that the spins point along this direct, provides the best agreement between measured and calculating splitting (see Fig. 4e of the Letter).

## Notes